\documentclass{PoS}

\newcommand{\be}{\begin{eqnarray}}
\newcommand{\ee}{\end{eqnarray}}

\title{Probing TMDs in heavy quarkonium production in $pp$ collision}

\ShortTitle{Probing TMDs}

\author{\speaker{Asmita Mukherjee}\thanks{Email :asmita@phy.iitb.ac.in}
~~and Sangem Rajesh\\
        Department of Physics, Indian Institute of Technology Bombay\\
        Powai, Mumbai 400076, India}

\abstract{We present a recent calculation of $J/\psi$ and $\Upsilon$
production in unpolarized $pp$ collision and show that this can be used to
probe the unpolarized gluon as well as the linearly polarized gluon 
transverse momentum dependent parton distributions (TMDs). We use the color
evaporation model for the heavy quarkonium production and use a generalized
factorized form of the cross section. We compare the results with
experimental data.}

\FullConference{QCD Evolution 2016\\
		May 30-June 03, 2016\\
		National Institute for Subatomic Physics (Nikhef), Amsterdam}

\begin{document}

\section{Introduction}

Single spin asymmetries (SSAs) when either the target or one of the
colliding protons is polarized have been experimentally observed since a
long time \cite{ssa1,ssa2,ssa3,ssa4,ssa5}.  Two main approaches to explain it theoretically are (1)
collinear framework \cite{collinear} involving higher twist quark or gluon correlators
 and (2) 
transverse momentum dependent distribution (TMD) and fragmentation functions \cite{jcollins}. 
While
the former approach is the first one and is free from several
complications associated with the TMD framework, the TMD based approach is
quite useful for phenomenological studies. Here one uses a generalized
factorized framework in terms of the TMDs. Gauge invariance needs the
inclusion of gauge links or Wilson lines in the operator definition of TMDs.
As these gauge links depend on the process in which the TMDs are probed,
this introduces process dependence. Thus there are issues related to
universality and applicability of factorization for different processes in
the TMD formalism. 
For simpler processes, like the semi-inclusive deep inelastic scattering
(SIDIS) or Drell-Yan (DY) the cross section can be written using a
generalized factorization. The TMDs are functions of the longitudinal momentum
fraction $x$ and transverse momentum $k_\perp$ of the partons (quark,
antiquark or gluon). These TMDs generate some asymmetries in the azimuthal
angle of the observed particle in the final state that can give information
on the spin and orbital angular momentum of the quarks and gluons. However,
it is also important to understand the unpolarized TMDs, not only because
they give the momentum distribution of the partons but also because they
appear in the denominator of the spin asymmetries, so one needs a good
understanding of them in order to understand the asymmetries \cite{melis}. 
$pp$
collisions are direct tools to probe the gluon TMDs, which play an important
role in the cross section and asymmetries in the collider kinematics. It is
known that there is a non-zero probability of finding linearly polarized
gluons in an unpolarized proton, provided they have non-zero transverse
momenta \cite{lin}. The corresponding TMD is denoted by  $h^{\perp g}_1$ and it is a
time-reversal even object. At leading twist, the gluon correlator of an
unpolarized proton is parametrized in terms of the unpolarized TMD
$f_1$ and the linearly polarized gluon TMD  $h^{\perp g}_1$. Although                      
 $h^{\perp g}_1$ has not been extracted from data yet, there are already
quite a few theoretical studies for possible extraction from different
experiments \cite{bm1,bm2,bm3,bm4,bm5,bm6,bm7,bm8}. Here we present a recent study of the possibility to probe it
in heavy quarkonium production in $pp$ collision \cite{us}.

\section{CHARMONIUM ($J/\psi$) AND BOTTOMONIUM ($\Upsilon$)
PRODUCTION CROSS SECTION }\label{sec2}  

There are mainly three models for heavy quarkonium production. In all these
models, the cross section is factorized into a hard part where the quarks
and gluons form the heavy quark and antiquark pair, and a soft or
non-perturbative part where the heavy quark pair   
forms a bound state with
definite quantum numbers. In the color singlet model (CSM) the heavy quark
pair is formed in a color singlet state. In the non-relativistic QCD (NRQCD)
based approach the heavy quark pair can be produced in  both color octet and
color singlet state, and the long distance factor or the non-perturbative
matrix element for the formation of the quarkonium can be expanded  in powers
of $v$ where $v$ is the relative velocity of the heavy quark in the
quarkonium rest frame. In the color evaporation model (CEM) \cite{cem}, that we use in
this calculation, the heavy quark pair radiates soft gluons to form a
quarkonium state of definite quantum numbers. Color of the $Q {\bar{Q}}$
pair does not affect the color of the bound state. The long distance factors 
in this model are considered to be independent of the process and obtained
by fitting data. The cross section for charmonium production in CEM is given
by \cite{us_sivers} :
\be
\sigma=\frac{\rho}{9}\int_{2m_Q}^{2m_{Q\overline{q}}}dM\frac{d\sigma_{Q\overline{Q}}}{dM},
\ee
where $m_Q$ is the mass of the heavy quark and $m_{Q\overline{q}}$ is the mass of lightest
heavy meson.  $M$ is the invariant mass of the  $Q\overline{Q}$ pair.
 $\rho$ is long distance factor and we took  0.47  and 0.62
for production of $J/\psi$ and $\Upsilon$ respectively.

We consider unpolarized proton-proton collision
\be \label{eqp}
h(P_A)+h(P_B)\rightarrow Q\overline{Q}(q)+X,
\ee
where the four momenta of the particles are given within round brackets.
The leading order (LO) subprocesses are $gg \rightarrow Q{\bar{Q}}$ and
$q{\bar{q}} \rightarrow Q{\bar{Q}}$. The differential cross section assuming
generalized factorization is written as :

\be\label{crosssection}
 \frac{d^4\sigma}{dydM^2d^2{\bf q}_{T}}&={}&\frac{\rho}{18}\int dx_{a} dx_{b}  
 d^{2}{\bf k}_{\perp a} d^2{\bf k}_{\perp b}\delta^4(p_{a}+p_{b}-q)
 \Phi^{\mu\nu}_g(x_{a},{\bf k}_{\perp a})\Phi_{g\mu\nu}(x_{b},{\bf k}_{\perp b})
  {\hat{\sigma}^{g g\rightarrow Q\bar{Q}}}.\nonumber \\
\ee
 
Here $q_T$ is the transverse momentum of the quarkonium and in the
center-of-mass frame of the incident hadrons, where each of the hadrons move
along the $z$ axis. $k_{\perp a}$ and $k_{\perp b}$  are the transverse
momenta of the incoming gluons; $\Phi^{\mu \nu}_g$ is the gluon  correlator which are parametrized in term
soft TMDs. Contribution from the $q{\bar{q}}$ channel is found to be very
small in the kinematics of the colliders considered. So we consider only the
$gg$ channel. At leading twist, the parametrization of the gluon correlator
is given by,

\be\label{gcorr}
\Phi^{\mu\nu}_g(x,{\bf k}_{\perp})&=&\frac{n_\rho n_\sigma}{({\textit {k}}.n)^2}\int 
\frac{d(\lambda .{\textit{ P}})d^2\lambda_T}{(2\pi)^3}
e^{ik.\lambda}\langle P|\mathrm{Tr}[F^{\mu\rho}(0)F^{\nu\sigma}(\lambda)]|P\rangle|_{LF}\\
=&&-\frac{1}{2x}\left\{g^{\mu\nu}_Tf^g_1(x,{\bf k}_{\perp}^2)-\left(\frac{k^{\mu}_{\perp}k^{\nu}_{\perp}}
{M^2_h}+g^{\mu\nu}_T\frac{{\bf k}^2_{\perp}}{2M_h^2}\right)
h^{\perp g}_1(x,{\bf k}_{\perp}^2)\right\}.
\ee

$k^2_{\perp}=-{\bf k}^2_{\perp}$, $g^{\mu\nu}_T=g^{\mu\nu}
-P^{\mu}n^{\nu}/P.n-n^{\mu}P^{\nu}/P.n$ and $M_h$ is the mass of 
proton. The unpolarized and the linearly polarized gluon
distribution functions are denoted by $f^g_1(x,{\bf k}_{\perp}^2)$ and
$h^{\perp g}_1(x,{\bf k}_{\perp}^2)$, respectively. In terms of the TMDs,
the differential cross section takes the form :

\be
\frac{d^4\sigma}{dydM^2d^2{\bf q}_{T}}&=&{} \frac{\rho}{18}\int
\frac{dx_{a}}{2x_a}\frac{dx_{b}}{x_b}d^{2}{\bf k}_{\perp a} d^{2}{\bf k}_{\perp
b}
\delta^4(p_{a}+p_{b}-q)\Big\{f_1^g(x_{a},{\bf k}_{\perp a}^2)f_1^g(x_{b},{\bf k}_{\perp b}^2)\nonumber \\
&&~~~~~~~~~~ +wh_1^{\perp g}(x_{a},{\bf k}_{\perp a}^2)h_1^{\perp g}(x_{b},{\bf
k}_{\perp b}^2)\Big\}
\hat{\sigma}^{g g\rightarrow Q\overline{Q}}(M^2)
\ee
where $w$ is weight factor:
\be
w=\frac{1}{2M_h^4}\left[({\bf k}_{\perp a}.{\bf k}_{\perp b})^2-\frac{1}{2}
{\bf k}_{\perp a}^2{\bf k}_{\perp b}^2\right].
\ee
As stated above, we neglect the contribution from the $q{\bar{q}}$ channel. Cross section for
the gluon initiated subprocess is calculated perturbatively. Using the
momentum conserving delta function, we obtain
\be
x_{a,b}=\frac{M}{\sqrt{s}}e^{\pm y},
\ee
where $y$ is the rapidity and $\sqrt{s}$ is the center-of-mass energy of the experiment.

\section{MODEL FOR TMDS AND TMD EVOLUTION}

We assume a Gaussian form for  the transverse momentum dependence of the TMDs
\cite{tmdmodel} : 

\be
 f_1^{g}(x,{\bf k}^2_{\perp })=f_1^{g}(x,Q^2)\frac{1}{\pi \langle
k^2_{\perp }\rangle}
 e^{-{\bf k}^2_{\perp }/\langle k^2_{\perp }\rangle}.
\ee

$f_1^{g}(x,Q^2)$ is the unpolarized gluon distributions (pdfs), the
scale is given by $Q^2=M^2$. For the numerical calculation, we have chosen
MSTW2008 distribution \cite{mstw}.  The
factorized form of $h_1^{\perp g}$ \cite{bm8} is given by
\be\label{hg}
h_1^{\perp g}(x,{\bf k}^2_{\perp })=\frac{M^2_hf_1^g(x,Q^2)}{\pi\langle
k^2_{\perp }\rangle^2}\frac{2(1-r)}{r}e^{1-
 {\bf k}^2_{\perp }\frac{1}{r\langle k^2_{\perp }\rangle}},
\ee
where $r$ is the parameter which has the range $0<r<1$. We have chosen two
values for $r$, $r=1/3$ and $r=2/3$. We use two values for squared intrinsic
average transverse momentum of gluons and quarks: 
$\langle k^2_{\perp }\rangle=0.25$ GeV$^2$ and 1 GeV$^2$ \cite{bm8}.
In model I, we have integrated over the full range of $k_\perp$ whereas in
model II, we have used an upper bound,  $k_{\mathrm{max}}=\sqrt{\langle
{k}^2_{\perp a}\rangle}$. Evolution of the unpolarized pdfs with the scale
is given by the DGLAP evolution equation. On the other hand, TMD 
evolution is performed in $b$ space \cite{tmde3}. In terms of the TMDs in $b$ space, the
differential cross section is given by :
\be
 \frac{d^4\sigma}{dydM^2d^2{\bf q}_T}&={}&\frac{\rho}{18s}\frac{1}{2\pi}
 \int_0^{\infty}b_{\perp} db_{\perp}J_0(q_Tb_{\perp})
 \Big\{ f_1^g(x_{a}, b_{\perp}^2)f_1^g(x_{b}, b_{\perp}^2)\\ 
&& +h_1^{\perp g}(x_{a}, b_{\perp}^2)h_1^{\perp g}(x_{b},b_{\perp}^2)\Big\}
  {\hat{\sigma}^{g g\rightarrow Q\overline{Q}}(M^2)},
\ee     
where $J_0$ is the Bessel function. The scale dependence of the TMDs is not
explicitly shown above. They depend on the renormalization scale $\mu$ and
the auxiliary parameter $\zeta$. Using Collin-Soper and renormalization
group equations, we can write \cite{tmde3}:
\be
  f(x,b_{\perp},Q_f,\zeta)=f(x,b_\perp,Q_i,\zeta)R_{pert}\left(Q_f,Q_i,b_{\ast}\right)
  R_{NP}\left(Q_f,Q_i,b_{\perp}\right),
 \ee
where $R_{pert}$ and $R_{NP}$ denote the perturbative and non-perturbative parts
 of the evolution kernel, respectively. $c/b_{\ast}$
 is the initial scale where $c=2e^{-\gamma_\epsilon}$ with the Euler's 
 constant $\gamma_\epsilon\approx0.577$. We have used the $b_{\ast}$
 prescription, with 
 $b_{\ast}(b_{\perp})=
\frac{b_{\perp}}{\sqrt{1+\left(\frac{b_{\perp}}{b_{\mathrm{max}}}\right)^2}}\approx  b_{\mathrm{max}}$.
 We have used the leading order (LO) anomalous dimensions in $R_{pert}$ and
for $R_{NP}$, the parametrization from \cite{tmde1}. No experimental data
is yet available for the extraction of $h_1^{\perp g}$, and we use the same
$R_{NP}$ for it as for the unpolarized distribution.

\section{NUMERICAL RESULTS}

We calculate the transverse momentum distribution for $J/\psi$
and $\Upsilon$ production. For $J/\psi$ production, 
we took the  charm quark mass ($m_c=1.275$ GeV) for $m_Q$ and lightest D meson mass ($m_D=1.863$ GeV) for $m_{Q\bar{q}}$. 
For the $\Upsilon$, bottom quark mass ($m_b=4.18$ GeV) for  $m_Q$ and lightest B meson mass ($m_B=5.279$ GeV) for $m_{Q\bar{q}}$
were used. The ranges of rapidity integration are : $y\in[2.0,4.5]$, $y\in[-3.0,3.0]$ and 
 $y\in[-0.5,0.5]$ for LHCb, RHIC and AFTER respectively, to obtain the differential 
cross section as a function of $q_T$.

\begin{figure}[t]
\begin{minipage}{0.99\textwidth}
\includegraphics[width=7.5cm,height=6cm]{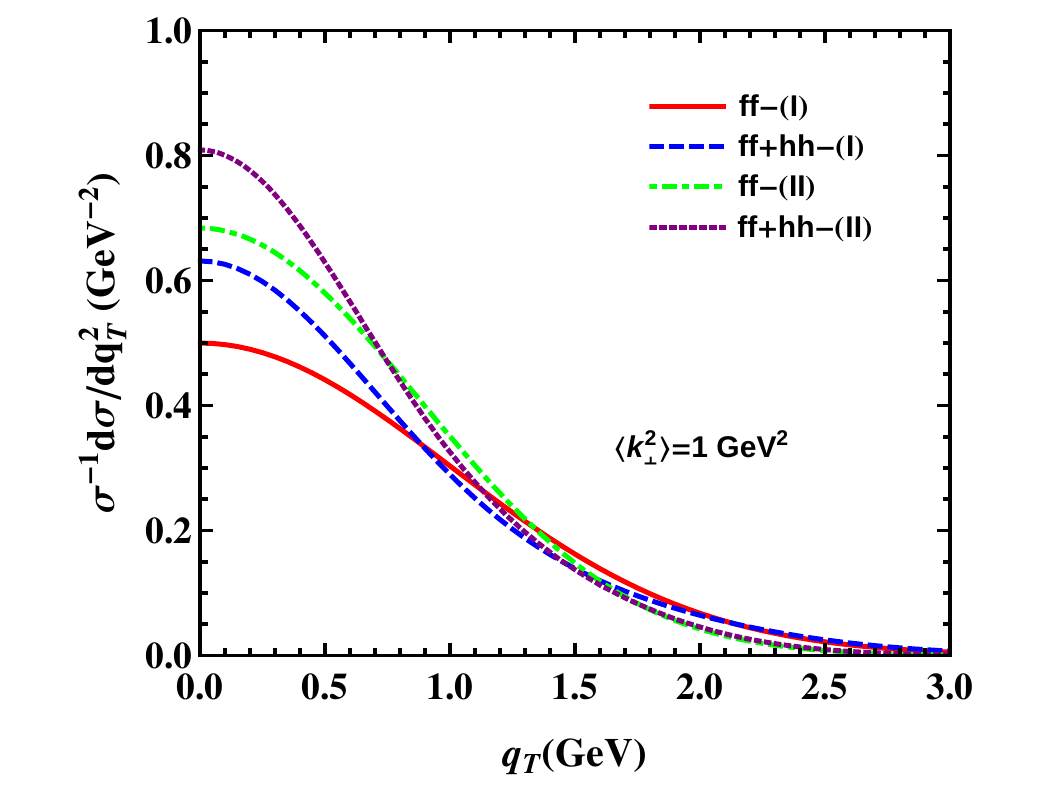}
\hspace{0.1cm}
\includegraphics[width=7.5cm,height=6cm]{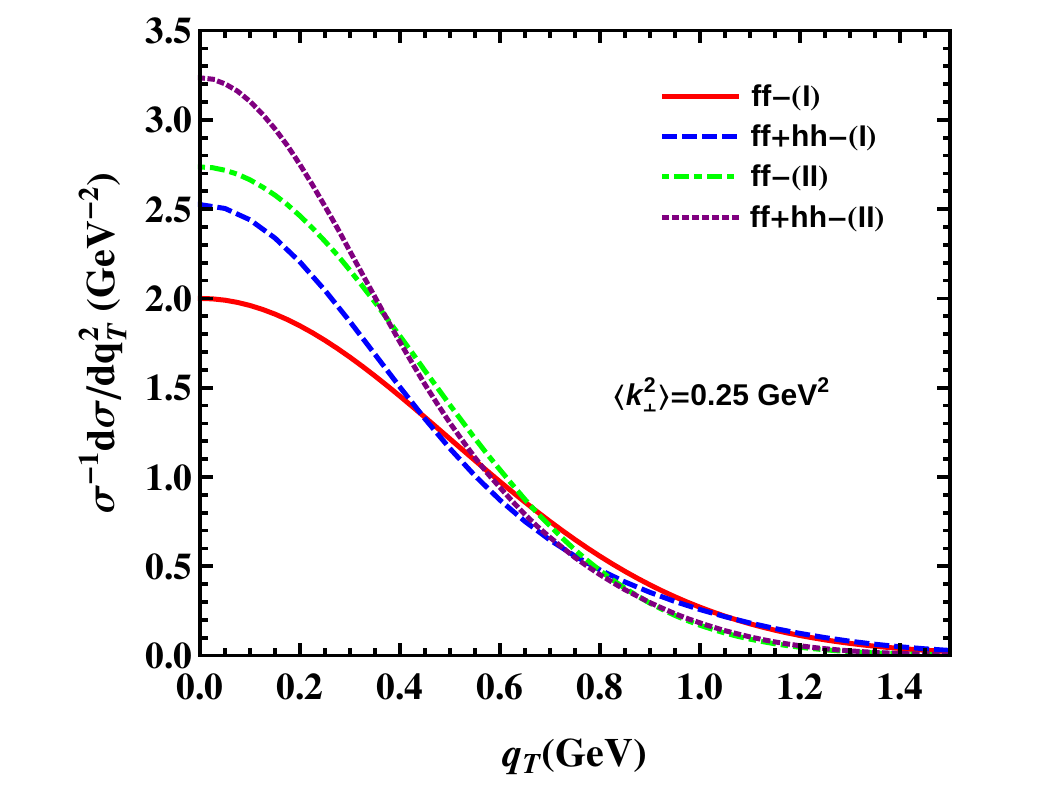}
\end{minipage}
\caption{\label{fig1} Normalized differential cross section of 
 $J/\psi$  and $\Upsilon$ production in ${pp} \rightarrow Q{\bar{Q}}+X$  
at LHCb ($\sqrt{s}=7$ TeV),  RHIC ($\sqrt{s}=500$ GeV) and AFTER ($\sqrt{s}=115$ GeV) energies
using DGLAP evolution approach  for  $r=\frac{2}{3}$ .
The solid (ff-(I)) and dot dashed (ff-(II)) lines are obtained by considering
unpolarized gluons and quarks in Model-I and Model-II respectively.
The dashed (ff+hh-(I)) and tiny dashed (ff+hh-(II)) lines are obtained by  taking 
into account unpolarized gluons and quarks plus linearly polarized gluons in Model-I and Model-II
respectively \cite{us}. }
\end{figure}

The $q_T$ distributions for $J/\Psi$ and $\Upsilon$ at the center-of-mass
energies of different experiments are shown in Fig. \ref{fig1}. We have 
normalized the results by    
the total cross section. In this plot, we have not incorporated the TMD
evolution, instead only the DGLAP evolution of the unpolarized pdf is used.
The normalized results  overlap for the different kinematics of different
experiments. The results are larger in magnitude in model II compared to
model I. In particular for lower values of $q_T$, the effect of linearly
polarized gluons are seen in the cross section. Above $q_T \approx 1~~
\mathrm{GeV}$ this effect is not seen any more. We have shown the results
for two values of  $\langle k^2_{\perp}\rangle$, namely $0.25$ and 
$1$~ $\mathrm{GeV}^2$ respectively. For small value of the Gaussian width,
the magnitude is higher. We have chosen $r=2/3$.   

\begin{figure}[t]
\begin{minipage}[c]{0.99\textwidth}
\tiny{(a)}\includegraphics[width=7.5cm,height=6cm,clip]{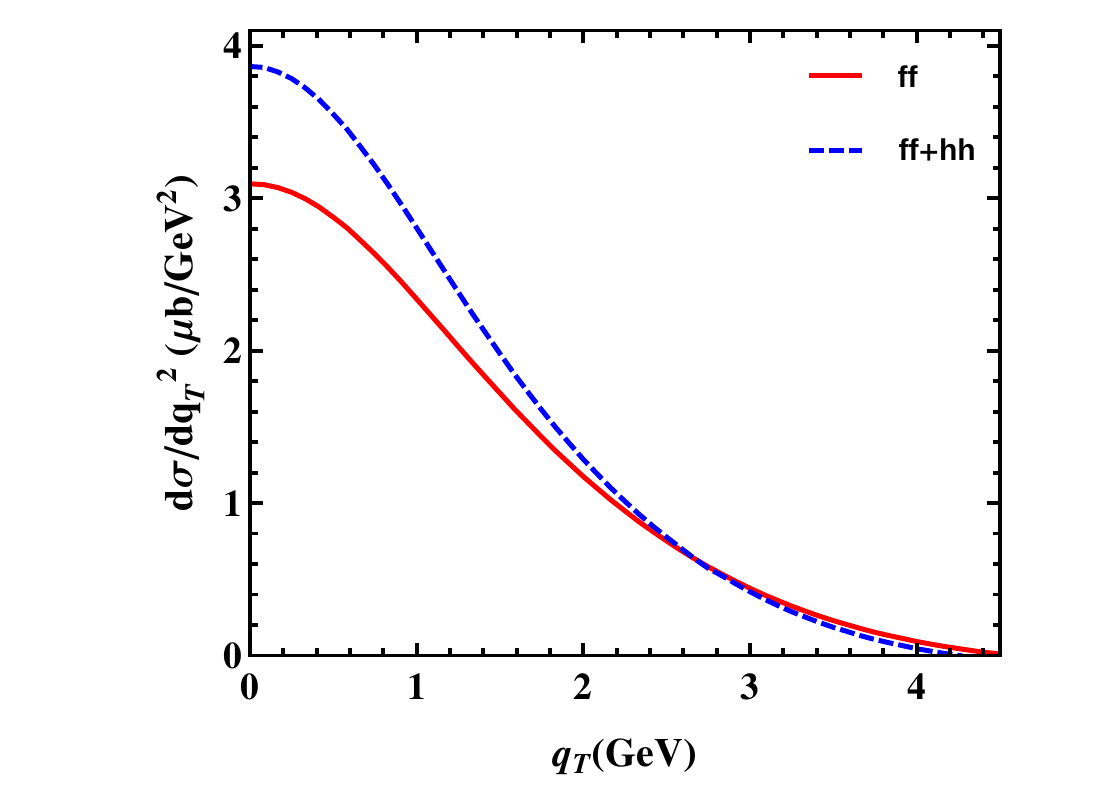}
\hspace{0.1cm}
\tiny{(b)}\includegraphics[width=7.5cm,height=6cm,clip]{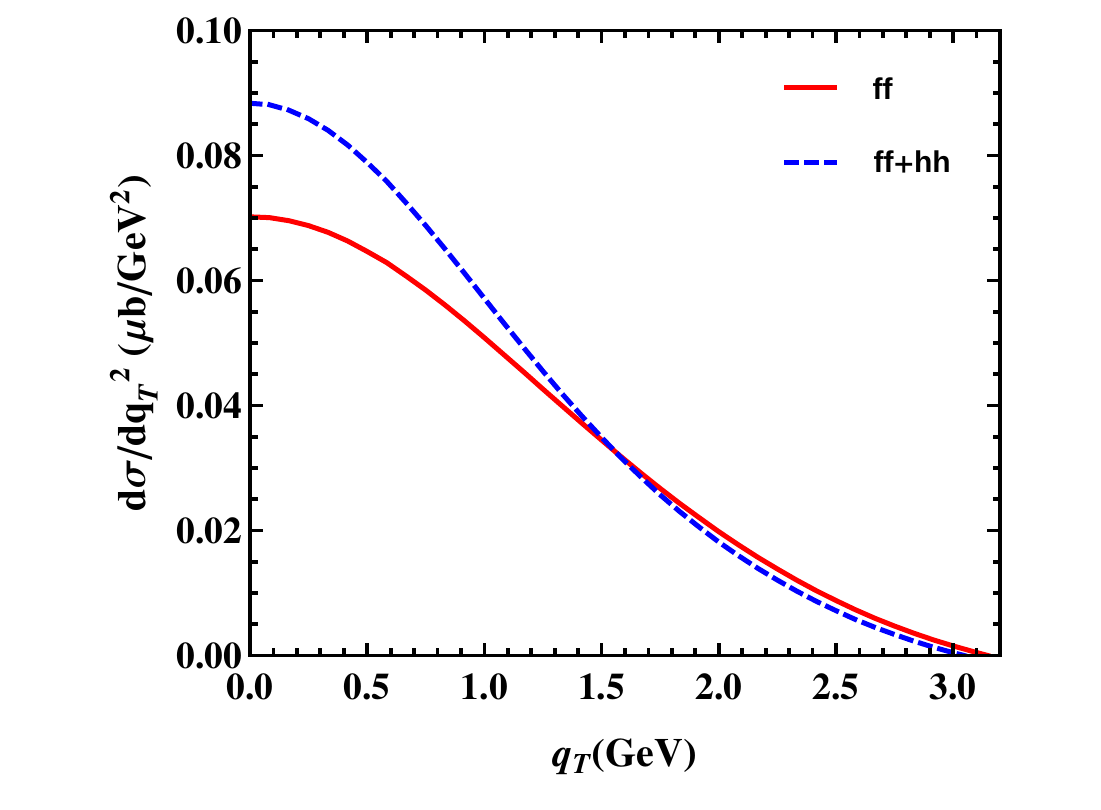}
\end{minipage}
\caption{\label{fig2}Differential cross section of  $J/\psi$ 
 as function of $q_T$  at LHCb ($\sqrt{s}=7$ TeV)  (left panel) and AFTER ($\sqrt{s}=115$ GeV)
(right panel) energies  using TMD evolution approach.  The solid (ff) and dashed (ff+hh)  lines are obtained
by considering  unpolarized gluons only and unpolarized plus linearly polarized gluons 
respectively \cite{us}.}
\end{figure}

In Fig. \ref{fig2}, we have shown the  $q_T$ distributions for $J/\Psi$
production for the kinematics of LHCb and AFTER at LHC respectively. 
In these plots we have incorporated the TMD evolution. The results are not
normalized here by the total cross section. We see again that at low $q_T$
values the cross section is modified when contribution from linearly
polarized gluons are taken into account.

\begin{figure}[t]
\begin{minipage}[c]{0.99\textwidth}
\includegraphics[width=7.5cm,height=6cm,clip]{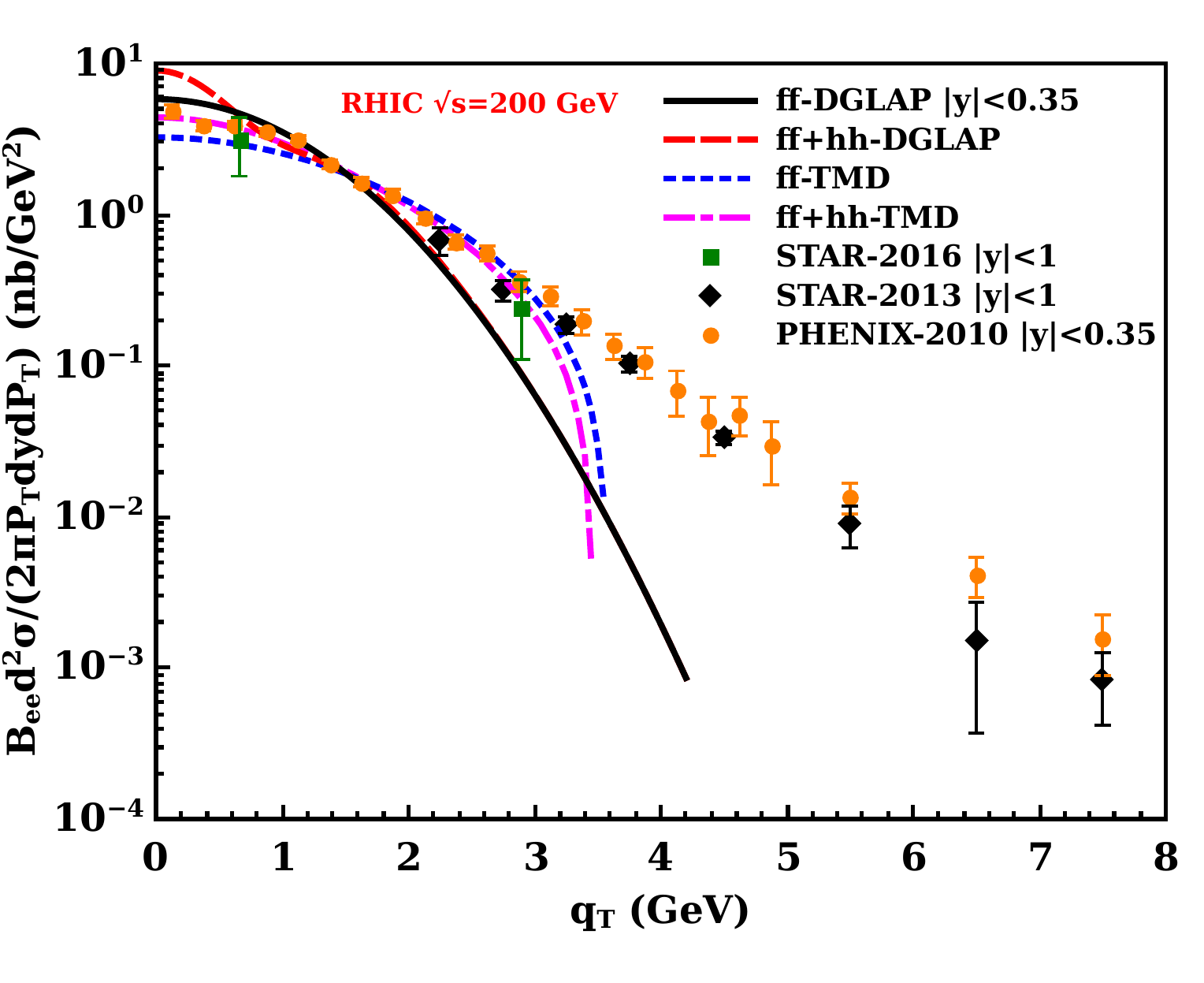}
\end{minipage}
\caption{\label{fig3} Differential cross section of  $J/\psi$  production
calculated in CEM model as function of transverse momentum  in the dielectron decay
channel. Center-of-mass energy is $200$ GeV. DGLAP denotes results calculated using DGLAP evolution for the
unpolarized pdf, TMD indicates the results are calculated in the TMD evolution
approach.  Theoretical results are compared with experimental data from the STAR
\cite{star,star_old} and PHENIX \cite{phenix} experiment at RHIC.} 
\end{figure}

In Fig. \ref{fig3}, we have compared our results with the experimental data
from STAR \cite{star, star_old} and PHENIX experiments \cite{phenix}  at
RHIC. Here we use the dielectron decay channel of $J/\Psi$. $B_{ee}$ is the
branching ratio for this channel. In this plot we have used the overall
normalization to be $\rho=0.9$. It is seen that the data is described well
by the theoretical plot, especially for low values of $q_T$. The TMD
evolved plots match the data upto $q_T \approx 3~$ GeV, and the plots using
DGLAP evolution fall faster and match the data only upto $q_T \approx 2 $ ~
GeV. The effect of the linearly polarized gluons in the cross section is not
that much visible, due to the log scale of the $y$-axis. It has been shown
that CEM explains the data quite well till about $q_T =10$~ GeV \cite{star}                
when higher order corrections are incorporated. Further work in this
direction in the TMD approach would include the process dependent gauge
links, and also the so called $Y$ -term, which we did not include in our
phenomenological study here.

\section {Conclusion}
 We presented a recent calculation of heavy quarkonium production in
unpolarized $pp$ collision in CEM using TMD formalism. At leading order the
gluon-gluon channel dominates. We have shown that the cross section has
substantial effect from the linearly polarized gluons at low $q_T$ of the
heavy quarkonium. We predicted the results for the kinematics of different
experiments and compared with data from RHIC. We found that the TMD evolution
formalism gives a better agreement with the data. Thus, heavy quarkonium
production in $pp$ collision is an important tool to probe the unpolarized
gluon TMDs and linearly polarized gluon TMDs.             

\section{Acknowledgement}

AM thanks the organizers of the QCD Evolution Workshop at Nikhef, for the
invitation and support.


\begin{thebibliography}{99}
\bibitem{ssa1} R. D. Klem, J. E. Bowers, H. W. Courant, H. Kagan, M. L. Marshak, E. A. Peterson, K. Ruddick,
W. H. Dragoset, and J. B. Roberts, Phys. Rev. Lett. 36, 929 (1976).
\bibitem{ssa2} G. Bunce et al., Phys. Rev. Lett. 36, 1113 (1976).
\bibitem{ssa3}D. L. Adams et al. (E704, E581), Phys. Lett. B261, 201 (1991).
\bibitem{ssa4}D. L. Adams et al. (FNAL-E704), Phys. Lett. B264, 462 (1991).
\bibitem{ssa5}I. Arsene et al. (BRAHMS), Phys. Rev. Lett. 101, 042001 (2008), 0801.1078.
\bibitem{collinear}K. Kanazawa, Y. Koike, A. Metz, D. Pitonyak, Phys. Rev D 91 (2015), 014013; Leonard Gam-
berg, Zhong-Bo Kang, Andreas Metz, Daniel Pitonyak, Alexei Prokudin, Phys.Rev. D 90 (2014),
074012.
\bibitem{jcollins}J. Collins, Foundations of Perturbative QCD, Cambridge University Press, 2011.

\bibitem{melis} S. Melis, EPJ Web Conf. {\bf 85},  01001 (2015).
\bibitem{lin} P. J. Mulders and J. Rodrigues, Phys. Rev. {\bf D63}, 094021
(2001).
\bibitem{bm1} D. Boer, S. J. Brodsky, P. J. Mulders, C. Pisano, Phys. Rev.
Lett. {\bf 106}, 132001 (2011).

\bibitem{bm2} J. W. Qiu, M. Schlegel, W. Vogelsang, Phys. Rev. Lett. {\bf
107}, 062001 (2011).

\bibitem{bm3} C. Pisano,  D. Boer, S. J. Brodky, M. G. A. Buffing, P. J.
Mulders, J. High. Energy. Phys. {\bf 10}, 024 (2013).

\bibitem{bm4}  D. Boer, P. J. Mulders, C. Pisano, Phys. Rev. {\bf D 80},
094017 (2009).

\bibitem{bm5} P. Sun, B. W. Xiao, F. Yuan, Phys. Rev. {\bf D 84}, 094005
(2011).

\bibitem{bm6} D. Boer, W. J. den Dunnen, C. Pisano, M. Schlegel  and W.
Vogelsang, Phys. Rev. Lett. {\bf 108}, 032002 (2012).

\bibitem{bm7} D. Boer, W. J. den Dunnen, C. Pisano, M. Schlegel,
 Phys. Rev. Lett. {\bf 111}, 032002 (2013).

\bibitem{bm8} D. Boer and C. Pisano, Phys. Rev. {\bf D 86}, 094007 (2012).

\bibitem{us} A. Mukherjee and S. Rajesh, Phys. Rev. {\bf D 93}, 054018
(2016). 

\bibitem{cem} F. Halzen, Phys. Lett. {\bf B 69}, 105 (1977), F. Halzen and
S. Matsuda, Phys. Rev. {\bf D 17}, 1344 (1978); H. Fritsch, Phys. Lett. {\bf
B 67}, 217 (1977). 
       
\bibitem{us_sivers} R. Godbole, A. Misra, A. Mukherjee, V. Rawoot, Phys. Rev. 
{\bf D 85}, 094013 (2012); Phys. Rev. {\bf D 88}, 014029 (2013).

\bibitem{tmdmodel} M. Anselmino, M. Boglione, U. D'Alesio, A. Kotzinian, F. Murgia,
 and A. Prokudin, Phys. Rev. D {\bf72}, 094007 (2005).

\bibitem{tmde3} S.M. Aybat, T.C. Rogers, Phys. Rev. D {\bf 83} (2011) 114042.

\bibitem{tmde1}  D. Boer and W. J. den Dunnen, Nucl. Phys. B {\bf 886}
(2014) 421.

\bibitem{star} L. Adamczyk {\it et al} (STAR Collaboration), Phys. Rev. {\bf C
93}, no.6, 064904 (2016).

\bibitem{star_old}  L. Adamczyk {\it et al} ( STAR Collaboration), Phys. Lett.
{\bf B 722}, 55 (2013).

\bibitem{phenix} A. Adare {\it et al} (PHENIX Collaboration), Phys. Rev. {\bf D
85}, 092004 (2012).   

\bibitem{mstw} A.D. Martin, W.J.Stirling, R.S. Thorne, G.Watt, 
Eur. Phys. J. C (2009) {\bf 63}: 189–285.
 
\end{thebibliography}

\end{document}